\documentclass[aip,jcp,superscriptaddress,reprint]{revtex4-1}
\usepackage{epsfig}
\usepackage{color}
\usepackage{amsmath}
\usepackage{amsfonts}
\usepackage{natbib}
\usepackage{notes2bib}
\usepackage[version=3]{mhchem} 

\newcommand{\avg}[1]{\left<#1\right>}
\newcommand{\len}[1]{\left|#1\right|}
\newcommand{\brac}[1]{\left[#1\right]}

\newcommand{\para}[1]{\left(#1\right)}

\newcommand{\ie}{\emph{i.e.}}
\newcommand{\eg}{\emph{e.g.}}

\newcommand{\etal}{\emph{et al.}}

\newcommand{\kT}{\ensuremath{k_{\rm B}T}}

\newcommand{\Zb}{\ensuremath{\mathcal{Z}}}

\begin{document}

\title{Exponential Scaling of Water Exchange Rates with Ion Interaction Strength
from the Perspective of Dynamic Facilitation Theory}

\author{Richard C. Remsing}
\email[]{rremsing@temple.edu}
\affiliation{Institute for Computational Molecular Science and Department of Chemistry,
Temple University, Philadelphia, PA 19122}

\author{Michael L. Klein}
\email[]{mike.klein@temple.edu}
\affiliation{Institute for Computational Molecular Science and Department of Chemistry,
Temple University, Philadelphia, PA 19122}

\begin{abstract}
Water exchange reactions around ionic solutes are ubiquitous in aqueous solution-phase chemistry.
However, the extreme sensitivity of exchange rates to perturbations in the chemistry of an ionic solute is
not well understood.
We examine water exchange around model ions within the language of dynamic facilitation theory,
typically used to describe glassy and other systems with collective, facilitated dynamics.
Through the development of a coarse-grained, kinetically-constrained lattice model of water exchange,
we show that the timescale for water exchange scales exponentially with the strength of the solute-solvent interactions. 
\end{abstract}

\maketitle

\raggedbottom

\section{Introduction}

Changes in coordination structures, particularly of coordinated water molecules,
are fundamental reactions of general importance in chemistry.
Understanding the molecular mechanisms underlying their chemical kinetics
is an important step in fully characterizing more complex reactions of broad importance.
For example, mineral dissolution and precipitation in aqueous solution
involves the coordination (and de-coordination) of metal ions and their complexes
by water molecules, and water exchange has been suggested to be a rate-limiting step in dilute solutions~\cite{AluniteDissolution,MineralRecrystallization,Casey_2007}.
In biological settings, water exchange reactions play an important role in the binding of ions to
RNA~\cite{Misra_2001,Elber_JPCB_2010,Klein_JACS_2010,Palermo_2015,Cunha01022017,Lee_2017}
and proteins~\cite{Kiriukhin:2002aa,Peraro_2007,Klein_JACS_2010,Palermo_2015,Stachura_2016},
influencing rates of biological processes.
Additionally, coordination complexes involving mixtures of water and a cosolvent
or pure non-aqueous solvents have proved important to the development of complex materials
for catalysis~\cite{Thenuwara_2016,Meyer:JPCC:2018}
and energy storage~\cite{Leung:JPCC:2013,Salanne_2016,Simoncelli_2018}.
The exchange of a hydration water for another around an ion (\ce{M}) --- the water exchange reaction ---
is one of the (seemingly) simplest changes in coordination structure,
\begin{equation}
\ce{M(OH_2)_n + ^*OH_2 <=> M(OH_2)_{n-1} ^*OH_2 + OH_2},
\end{equation}
where $\ce{^*OH_2}$ is used to identify a distinct water molecule,
and the ion \ce{M} has a coordination of \ce{n} in water.
The reactants and products are chemically identical in this process,
such that there is no net change in the free energy driving this reaction.
However, the kinetics governing water exchange are quite complex.
Typical timescales for water exchange depend on the nature of the ion
and vary from picoseconds to hundreds of years~\cite{WaterExchangeReview}.
In this work, we examine the complex dynamics of water exchange reactions within the framework
of dynamic facilitation (DF) theory, typically employed to describe glassy systems~\cite{Chandler_2010}.
 To do so, we suggest that the coordination of a solute by a water molecule imposes a kinetic constraint
 on the dynamics of that molecule. 
 We additionally suggest that these kinetic constraints are coupled to the charge density of the solute,
 and, in general, any solute-solvent interactions, such that
 strong solute-solvent interactions ultimately lead to slow, collective dynamics that are dominated by fluctuations,
 precisely the realm of applicability of DF theory.
 We use computer simulations to characterize water exchange around a range of model ions
 within the DF theory framework, illustrating that increased water exchange times
 correspond to increased glassiness and collective behavior in the ion hydration shell.
 Furthermore, by using DF theory and developing kinetically constrained lattice models,
 we predict that water exchange times scale exponentially with the strength of water-solute interactions,
 in agreement with recent empirical findings~\cite{Lee_2017}.
 %

\section{Simulation Details}

We probe the dynamics of water exchange around classical (Lennard-Jones plus point charge)
ion models using GROMACS version 5.0~\cite{gmx4ref}.
We do not consider the influence of changes in electronic structure~\cite{Parrinello_JCP_1999,Raugei_2002,Remsing:JPCL:2014,Klein_MolPhys_2015,Remsing_2018},
bond breaking~\cite{Hellstr_m_2017}, and nuclear quantum effects~\cite{Wilkins_2015},
which may impact water exchange kinetics.
All simulations were performed in the isothermal-isobaric (NPT) ensemble
at a temperature of $T=300$~K and a pressure of $1$~bar
using the canonical velocity rescaling thermostat of Bussi~\etal~\cite{Bussi:JCP:2007}
and an Andersen barostat~\cite{AndersenBaro}, respectively.
Water was modeled using the extended simple point charge (SPC/E) model~\cite{SPCE}.
We study several model ions to span a significant range of water exchange times:
(i) $\ce{Ca^{2+}}$, (ii) a model Cobalt ion with a reduced charge of 1.4 ($\ce{Co^{1.4+}}$) in order to make
direct sampling of water exchange possible, and (iii) an model potassium ion ($\ce{K^+}$).
$\ce{Co^{1.4+}}$ was modeled using the CM parameters developed by Merz and coworkers~\cite{Merz},
and the LJ parameters of the ECCR model were used for $\ce{Ca^{2+}}$,
with a charge of $+2$, neglecting any charge transfer between the ion and water~\cite{ECCR-Calcium}.
K$^+$ was modeled using the parameters developed by Koneshan~\etal~\cite{Koneshan:JPCB:1998,clayff}.
Short-ranged electrostatic and Lennard-Jones (LJ) interactions were truncated at a distance of 1~nm.
The particle mesh Ewald method was used to evaluate long-ranged electrostatic interactions~\cite{PME},
with the inclusion of a uniform neutralizing background potential
to cancel the net charge of the single ion in the system.
Equations of motion were integrated using a timestep of 0.5~fs, saving
configurations every 100 timesteps, employing the LINCS algorithm~\cite{PLINCS}
to fix the O-H bonds and H-O-H angles of water.
Trajectory lengths for $\ce{Co^{1.4+}}$, $\ce{Ca^{2+}}$, and $\ce{K^+}$ systems were
1~$\mu$s, 711~ns, and 100~ns, respectively.
Unless otherwise noted, results are presented for systems with 500 ($\ce{Co^{1.4+}}$ and $\ce{K^+}$)
or 512 ($\ce{Ca^{2+}}$) water molecules, which are large enough to minimize
finite size effects (see Supporting Information (SI)). 
%

\section{Facilitated Dynamics of Water Exchange}

\begin{figure}[tb]
\begin{center}
\includegraphics[width=0.47\textwidth]{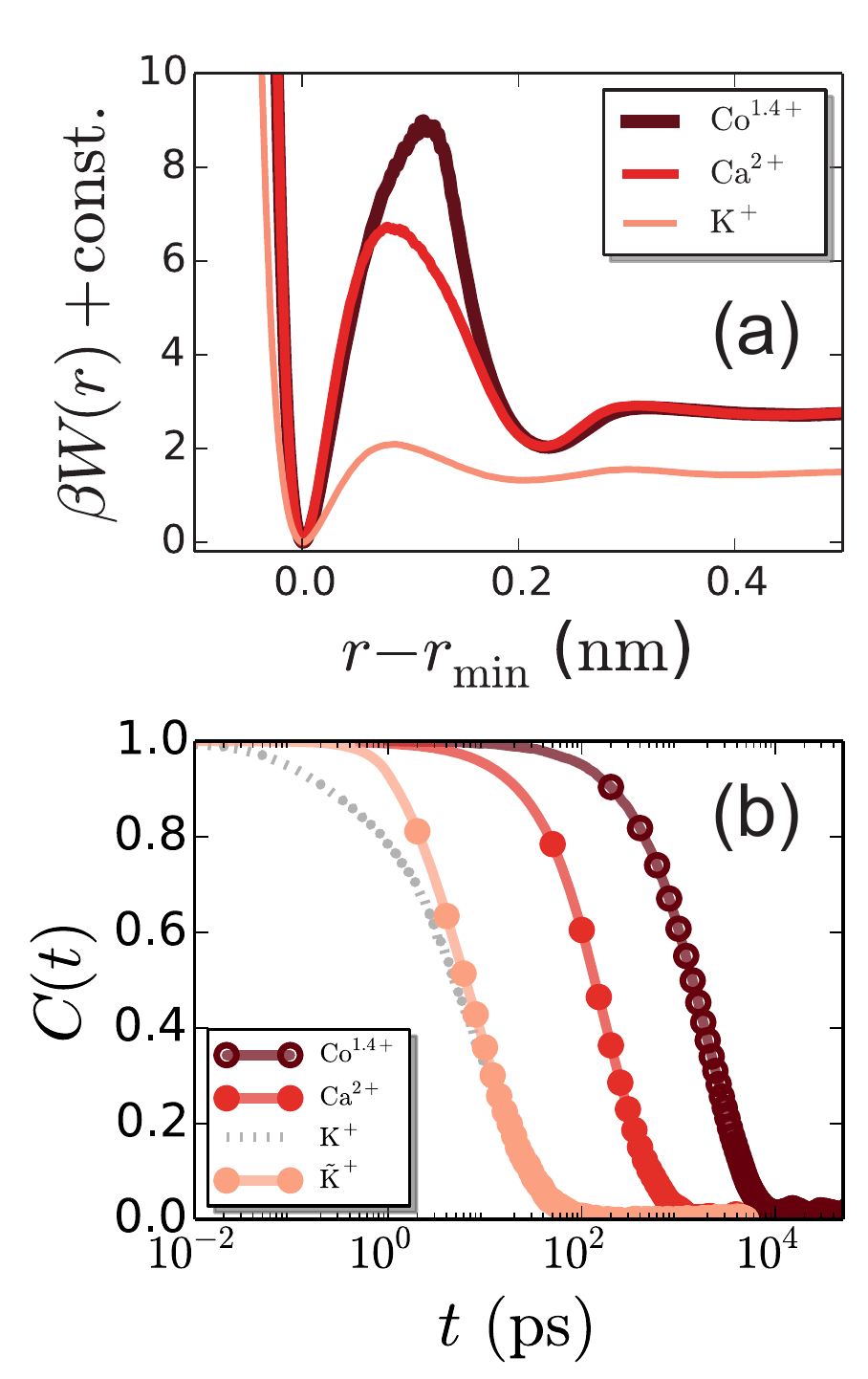}
\end{center}
\caption
{
(a) Ion-water potentials of mean force (PMFs), $\beta W(r) = -\ln g(r)$, for the three ions under study. 
Note that the PMFs have been shifted vertically and horizontally to place the free energy minimum at the origin.
(b) The time correlation function $C(t)$ for water exchange around
a reduced charge model of a cobalt ion (Co$^{1.4+}$),
a calcium ion (Ca$^{2+}$),
and a potassium ion (K$^+$).
For K$^+$, $C(t)$ is shown
with ($\tilde{\rm K}^+$) and without (dashed) coarse-graining in time over a window of 2~ps.
A similar coarse-graining does not impact $C(t)$ for the more highly charged ions.
}
\label{fig:rdf}
\end{figure}

In order to quantify the kinetics of water exchange, we must first define what we
mean by the exchange of a water molecule. 
Following previous work, we monitor exchange through the indicator function $h_i(t)$, which is equal to one
if water $i$ is in the hydration shell of an ion and is equal to zero when it is not.
A water molecule is in the first coordination shell if the ion-water oxygen distance is less
than $r_{\rm c}$, defined as the position of the free energy barrier in the solute-water oxygen potential of mean force,
$\beta W(r) = -\ln g(r)$,
shown in Fig.~\ref{fig:rdf}a for the ions under study,
where $g(r)$ is the water oxygen-ion pair distribution function, which are shown and discussed in
the SI, along with the ion coordination structure.
The free energy barriers increase following K$^+<$ Ca$^{2+}<$ Co$^{1.4+}$,
and we therefore expect the water exchange times to follow a similar trend.
%
%

%
Water exchange times are traditionally measured by computing the time correlation function
\begin{equation}
C(t)=\frac{\avg{\delta h_i(0) \delta h_i(t)}}{\avg{\delta h_i^2(0)}},
\end{equation}
which is shown in Fig.~\ref{fig:rdf}b,
where $\delta h_i(t)=h_i(t)-\avg{h_i(t)}$ and implicit in the ensemble average is an average over all water molecules.
For tightly bound water molecules, as is the case for $\ce{Ca^{2+}}$ and $\ce{Co^{1.4+}}$ model ions,
the decay of $C(t)$ is exponential, yielding water exchange lifetimes of roughly 200~ps for $\ce{Ca^{2+}}$
and 2000~ps for the reduced charge $\ce{Co^{1.4+}}$ model,
both of which are in good agreement with earlier estimates for similar model ion parameters~\cite{Lee_2017}.
Note that water exchange dynamics can display significant finite size effects, as discussed in the SIs
for $\ce{Ca^{2+}}$.
For  $\ce{K^{+}}$, $C(t)$ decays exponentially only after an initial transient period. 
This transient short-time dynamics physically arises from high-frequency recrossing of the barrier at $r_{\rm c}$
and is removed upon coarse-graining in the time domain,
as is typically done for glassy systems~\cite{Chandler_2010,Garrahan_2003,Keys_2011}.
The coarse-grained $C(t)$ for $\ce{K^{+}}$ decays exponentially with a time constant of roughly 10~ps,
where the coarse-graining is performed over a time window of 2~ps.
Coarse-graining the correlation functions for $\ce{Ca^{2+}}$ and $\ce{Co^{1.4+}}$ on this time scale
does not alter their behavior.
%

\begin{figure*}[tb]
\begin{center}
\includegraphics[width=0.95\textwidth]{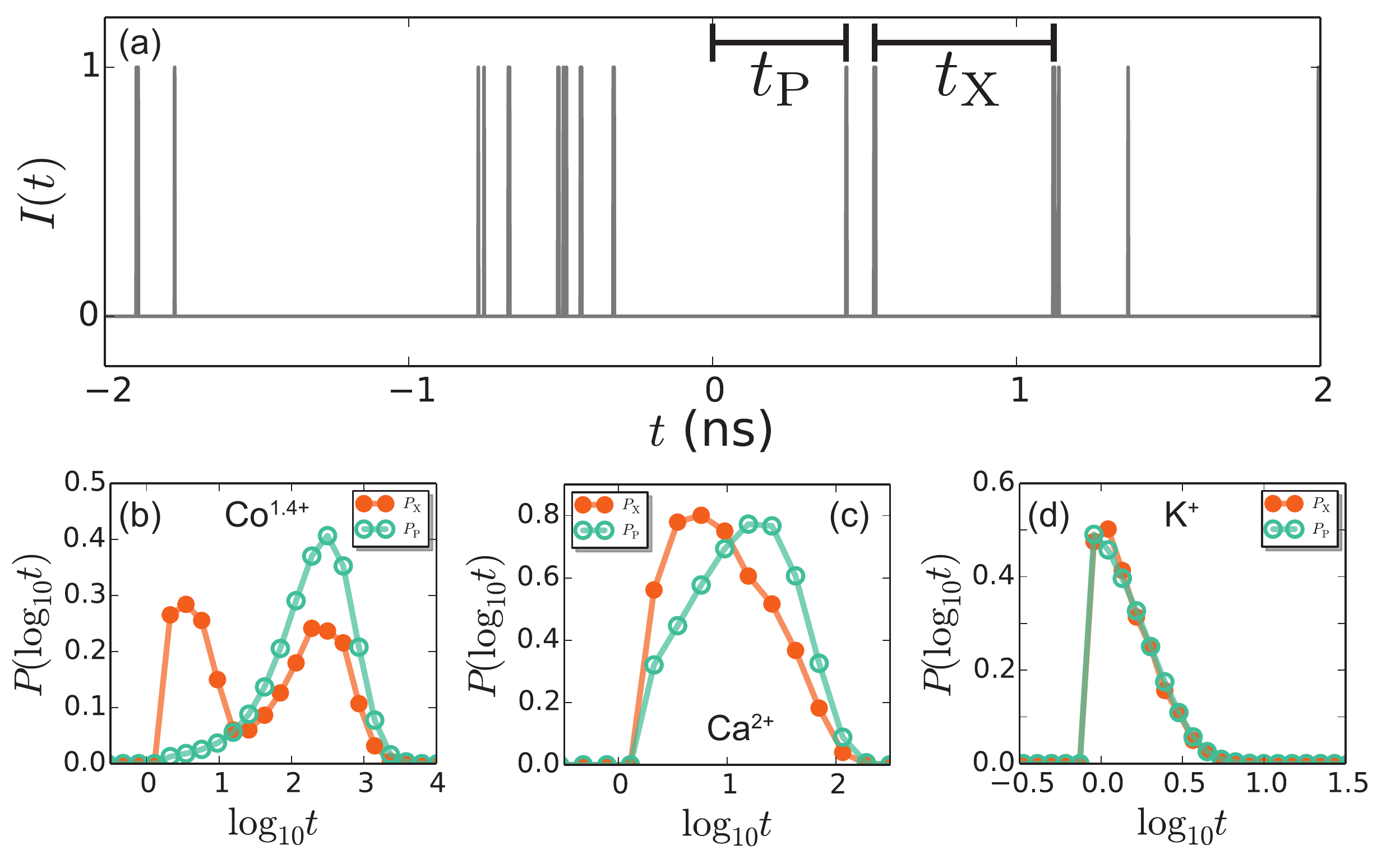}
\end{center}
\caption
{
(a) A portion of the $I(t)$ trajectory for Co$^{1.4+}$, highlighting individual persistence, $t_{\rm P}$,
and exchange time periods, $t_{\rm X}$, the averages of which are $\tau_{\rm P}=\avg{t_{\rm P}}$ and
$\tau_{\rm X}=\avg{t_{\rm X}}$, respectively.
(b-d) Distributions of the persistence and exchange times, $P_{\rm P}(t)$ and $P_{\rm X}(t)$, respectively,
for (b) Co$^{1.4+}$, (c) Ca$^{2+}$, and (d) K$^+$.
}
\label{fig:dist}
\end{figure*}

%
Exponential decay of correlation functions is typically indicative of uncorrelated dynamics,
not collective, heterogeneous dynamics.
However, we show below that the dynamics of water exchange are facilitated and collective, despite
the exponentially decaying $C(t)$.
Instead, we suggest that this exponential decay arises from the small phase space volume probed
by these correlation functions.
We characterize the facilitated nature of the dynamics of water exchange by computing the probability distributions
of \emph{exchange} and \emph{persistence} times, $P_{\rm X}(t)$ and $P_{\rm P}(t)$, respectively,
using the language of DF theory~\cite{Jung_2005,Hedges_2007}.
The exchange time is defined as the time between two successive events.
In contrast, the persistence time is the time for the first event to occur, as measured from $t=0$.
The distributions of exchange and persistence times are related through
\begin{equation}
P_{\rm P}(t) \sim \int_t^\infty dt' P_{\rm X}(t'),
\end{equation}
within a normalization constant~\cite{Jung_2005}.
We define an event as any change in the water coordination shell between two successive timesteps,
mathematically determined by the indicator function
\begin{equation}
I(t)=\Theta \para{ \Delta H(t) - c},
\end{equation}
where $\Theta(x)$ is the Heaviside step function, $c$ is a constant greater than zero,
$\Delta H(t)=\sum_{i=1}^N \Delta h_i(t)$,
and
$\Delta h_i(t) = \len{h_i(t)-h_i(t-\Delta t)}$.
The constant $c$ sets the threshhold for defining an event when the trajectory is coarse-grained in time
to avoid counting transient barrier recrossing, and here we choose $c=0.9$ without impact on the qualitative findings.
A representative portion of an $I(t)$ trajectory is shown in Fig.~\ref{fig:dist},
with individual exchange and persistence times highlighted.
This trajectory suggests that dynamics of water exchange are indeed facilitated.
One observes long periods of inactivity and bursts of activity that are localized in time.
This complex dynamic structure is quantified by the exchange and persistence time distributions.
For a Poisson process, the distribution of bound water lifetimes is exponential,
and the persistence and exchange time distributions coincide and are exponential~\cite{Jung_2005}. 
In the context of facilitated dynamics, a dynamical event cannot occur unless a molecule
is intersected by dynamical excitations in spacetime, which become increasingly sparse
under conditions where the dynamics are slowed.
In supercooled liquids, the concentration of excitations is typically controlled by the temperature.
In the context of water exchange, the relevant excitations are those that permeate the spacetime of the hydration shell,
and the density of these excitations is expected to be inversely related
to the solute-solvent interaction strength, in addition to the temperature of the solution.
In the limit of high solute charge density, for example, the excitations will be sparse, leading to persistence times
that are significantly decoupled from exchange times~\cite{Jung_2005,Hedges_2007}.
This behavior is exemplified by the exchange and persistence time distributions in Fig.~\ref{fig:dist}.
As the solute charge density is increased, spacetime excitations in the hydration shell become increasingly sparse,
and there is a small number of dynamic pathways that enable hydration water to exchange.
Consequently, the exchange and persistence time distributions become increasingly decoupled
as the solute-solvent interaction strength is increased, Fig.~\ref{fig:dist}b-c.
This illustrates that strong solute-solvent interactions lead to facilitated and collective water exchange processes.
For weak solute-solvent interactions (Fig.~\ref{fig:dist}d), as is the case for K$^+$,
water molecules in the hydration shell are not trapped by strong solute-solvent interactions,
and the barrier to water exchange is small (Fig.~\ref{fig:rdf}a). 
In this case, excitations permeate the spacetime of the solute hydration shell, there is a high density
of dynamic pathways that enable water exchange to occur, and the exchange and persistence
time distributions coincide.
We expect that solutes with net repulsions, including effective repulsions leading to drying around large
hydrophobic solutes~\cite{LCW,Chandler:Nature:2005,Remsing:JStatPhys:2011,Remsing:JPCB:2013,Remsing:PNAS:2016},
will also lead to $P_{\rm X}(t)$ and $P_{\rm P}(t)$ that coincide.
We also note that the exponential decay of $C(t)$ shown in Fig.~\ref{fig:rdf}b,
which seems to contradict the signatures of dynamic heterogeneity found here,
is expected to arise from the small spacetime volume in which the system is kinetically constrained.
We expect that a stretched exponential decay of $C(t)$ would be observed for large enough solutes,
as is found for strongly binding extended surfaces~\cite{Willard_2013,Limmer_2013,Limmer:CPL:2015}.
Similar finite size effects are well-documented in configuration~\cite{PhysRevB.34.1841,PhysRevB.30.1477,PhysRevLett.61.2635,PhysRevB.43.3265,PhysRevE.86.031502,PhysRevE.61.R41}
and trajectory~\cite{Chandler_2010,Merolle:PNAS:2005,Hedges:Science:2009} space.
%

\section{Kinetically Constrained Model of Solvent Exchange}

We now show that the general DF phenomenology of solvent exchange
reactions can be captured within the framework of kinetically constrained lattice models.
We illustrate this within the context of the Fredrickson-Andersen (FA) model~\cite{FA},
but our approach is general and can be readily extended to other kinetically constrained models, such as the East model~\cite{East}.
In the FA model, each lattice site is described by a binary variable $n_i=0,1$, typically interpreted as
(0) immobile and (1) mobile, or excited, sites.
The energy of a FA model with $N$ sites is $E=J\sum_{i=1}^N n_i$, where $J$ is an energy scale.
The resulting thermodynamic properties of the FA model are trivial.
In particular, the concentration of excitations (mobile sites) is 
$c_{\rm FA}=\avg{n_i} = 1/(1+e^{1/\tilde{T}})$,
where $\tilde{T}=\kT/J$.
The complex dynamics of the FA model arise from kinetic constraints imposed on the system.
The kinetic constraints dictate that a change in state of a site can only occur if at least one of its neighboring sites is excited,
with excitations ($0\rightarrow 1$) occurring at a rate of $c_{\rm FA}$
and de-excitations ($1\rightarrow 0$) occurring at a rate of $1-c_{\rm FA}$.
In the water exchange reaction, the solute imposes kinetic constraints largely on the water molecules in its coordination shell only,
not the rest of the solvent.
Thus, we introduce a fixed solute site into the lattice model, which, for simplicity, does not move or change state.
In order to mimic the water exchange reaction, we impose the FA kinetic constraints only on the nearest neighbor sites
of the solute; the remaining lattice sites exhibit unconstrained dynamics.
The solute is considered to be an immobile region; it does not facilitate the dynamics of its neighbors.
Finally, we note that the kinetics of the ionic solvation shell water molecules are altered by the nature of the solute,
through solute-solvent interactions.
To mimic this, we assign a different energy scale, $J'$, to the nearest neighbors of the solute in the lattice model, 
arising physically from any solute-solvent interactions present.
Defining $\tilde{T}'=\kT/J'$, one can consider the dynamics of these sites to evolve at an effective reduced temperature $\tilde{T}'$,
which is different than that of the rest of the system, $\tilde{T}$.
In this context, $\tilde{T}'$ captures the effects of the solute charge density and any other solute-solvent interactions (\eg \ LJ interactions).
For example, lowering $\tilde{T}'$ is the analog of increasing the charge density of the ion.
%

\begin{figure}[t]
\begin{center}
\includegraphics[width=0.47\textwidth]{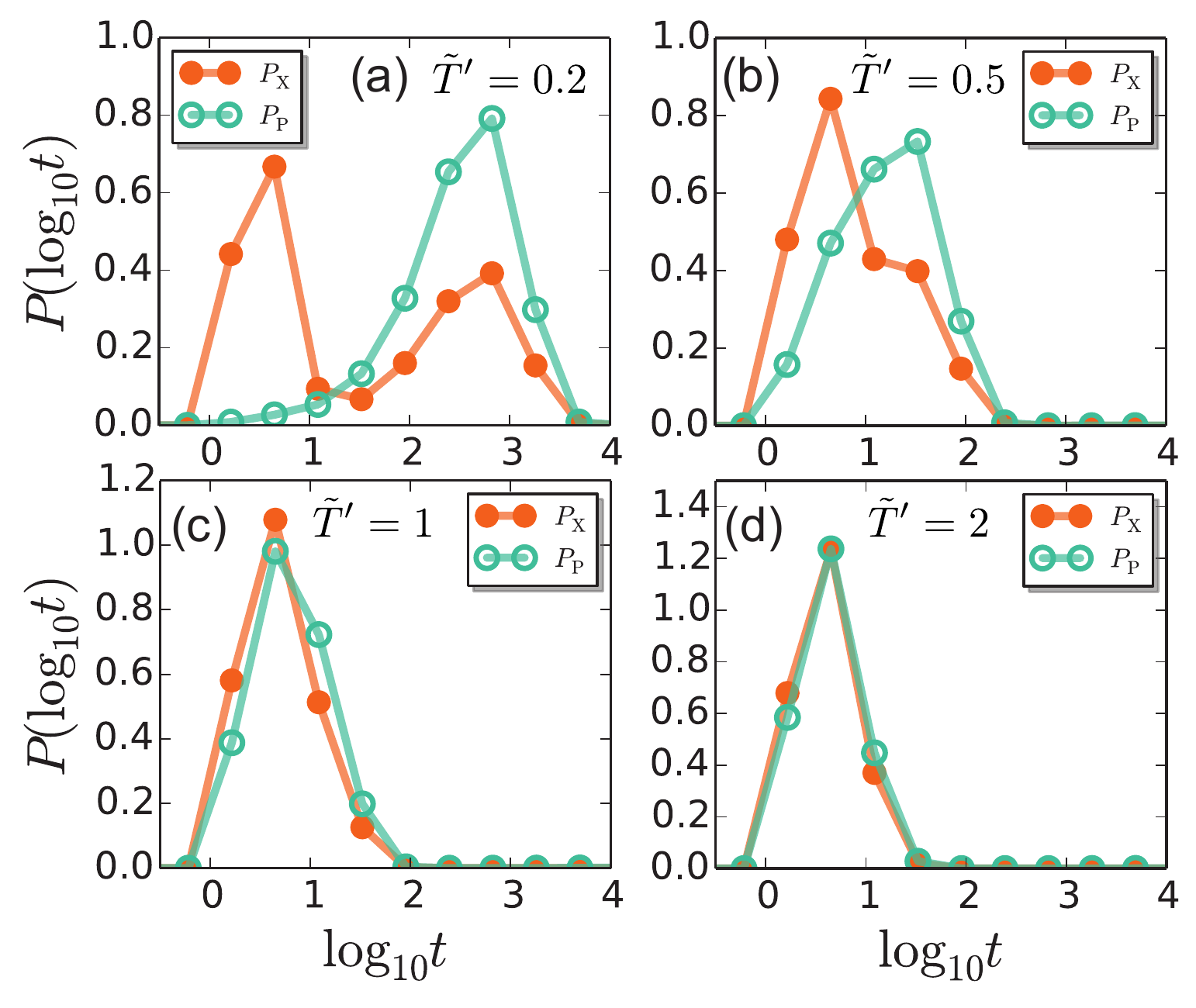}
\end{center}
\caption
{
Distributions of exchange and persistence times, $P_{\rm X}(t)$ and $P_{\rm P}(t)$, respectively,
for the kinetically constrained solvent exchange lattice model at solute nearest-neighbor effective temperatures
of (a) $\tilde{T}'=0.2$, (b) $\tilde{T}'=0.5$, (c) $\tilde{T}'=1$, and (d) $\tilde{T}'=2$.
Time is in units of Monte Carlo sweeps.
}
\label{fig:fadists}
\end{figure}

%
Here, we focus on the one-dimensional (1D) version of this model.
The thermodynamics of this model are trivial and analogous to that of the FA and East models.
For $S$ solutes, dispersed far apart (their solvation shells are not shared),
there will be $2S$ solute neighbors in a 1D system.
Omitting the (static) solutes from consideration, the (solvent) partition function is readily evaluated as
\begin{equation}
\Zb = (1+e^{-1/\tilde{T}})^{N-2S} (1+e^{-1/\tilde{T}'})^{2S}.
\end{equation}
The concentration of excitations is
obtained from the sum of the excitation concentrations in the constrained and unconstrained regions,
\begin{align}
c &= \para{1-\frac{2S}{N}}\frac{1}{1+e^{1/\tilde{T}}} + \frac{2S}{N}\frac{1}{1+e^{1/\tilde{T}'}} \nonumber \\
&\equiv \para{1-\frac{2S}{N}} c_{\rm B}+\frac{2S}{N} c_{\rm NN}.
\label{eq:ex}
\end{align}
The first term in Eq.~\ref{eq:ex}, $c_{\rm B}$, corresponds to the concentration of excitations in the unconstrained,
bulk (B) region of the system,
\ie \ the lattice sites that are not nearest neighbors of a solute.
Correspondingly, the second term in Eq.~\ref{eq:ex} is the concentration of excitations in the regions
composed of lattice sites that are nearest neighbors (NN) of a solute.
Thus, Eq.~\ref{eq:ex} is equal to $c_{\rm FA}$ when $S=0$ or $\tilde{T}'=\tilde{T}$.
Equation~\ref{eq:ex} illustrates that the role of the effective temperature $\tilde{T}'$
is to alter the concentration of excitations in the solute solvation shell.
Indeed, our interpretation of water exchange described in the previous section
yields a similar picture for the effects of solute-solvent interactions,
which are mimicked by $\tilde{T}'$.
The corresponding distributions of the exchange and persistence times are decoupled for small $\tilde{T}'$
(strong solute-solvent interactions), Fig.~\ref{fig:fadists}a-d,
which were obtained from 100 independent trajectories at each $\tilde{T}'$ using a 1D lattice with $N=50$ and $S=1$,
evolving under Metropolis dynamics.
As $\tilde{T}'$ is increased, the exchange and persistence time distributions become increasingly similar.
Physically, increasing $\tilde{T}'$ corresponds to weakening the solute-solvent interactions, and even
adding solute-solvent repulsions ($\tilde{T}'>\tilde{T}$), arising from direct or effective repulsions that can lead to drying around extended hydrophobic surfaces~\cite{LCW,Remsing:JPCB:2013,Remsing:PNAS:2016}. 
Both reduce the kinetic constraints for water exchange and lead to overlapping distributions.
%

\begin{figure}[tb]
\begin{center}
\includegraphics[width=0.42\textwidth]{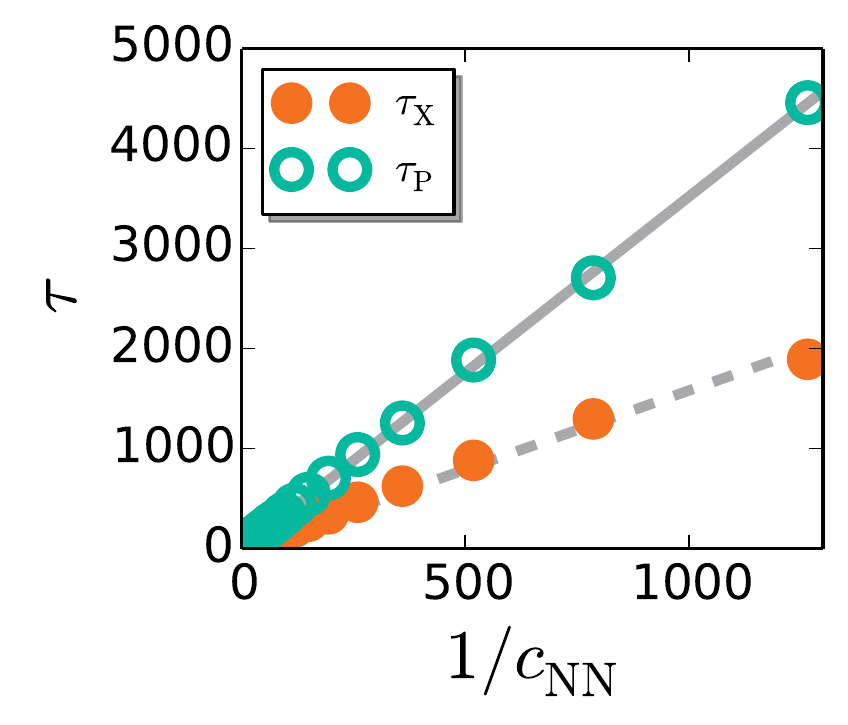}
\end{center}
\caption
{
Exchange and persistence times ($\tau_{\rm X}$ and $\tau_{\rm P}$, respectively)
as a function of the inverse concentration of excitations in the nearest-neighbor regions, $1/c_{\rm NN}$.
The dashed and solid lines correspond to linear scaling of $\tau_{\rm X}$ and $\tau_{\rm P}$, respectively, with $1/c_{\rm NN}$.
Time is in units of Monte Carlo sweeps.
Error bars are smaller than the symbol size.
}
\label{fig:avgfa}
\end{figure}

\begin{figure}[tb]
\begin{center}
\includegraphics[width=0.4\textwidth]{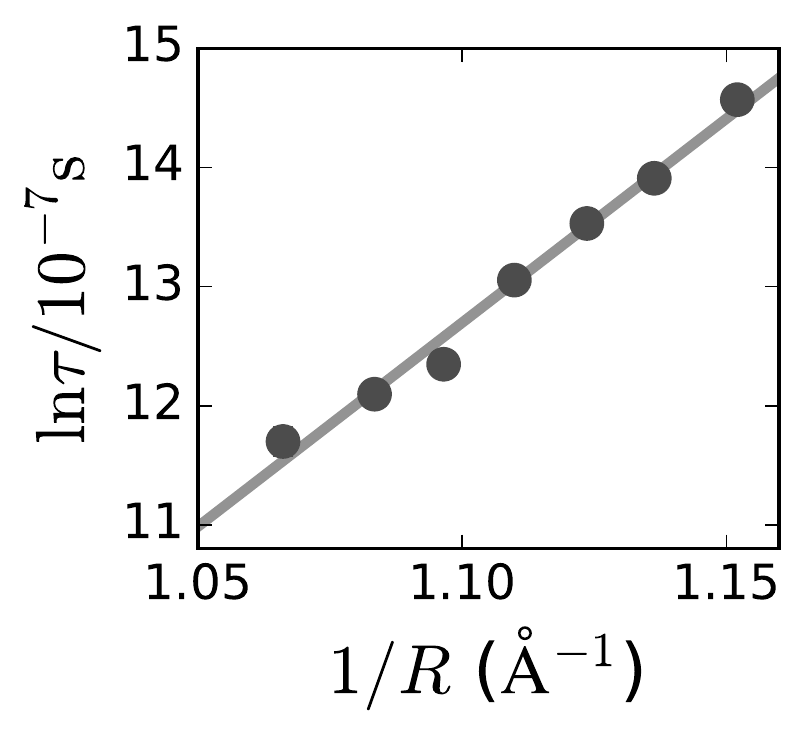}
\end{center}
\caption
{
Experimentally determined water exchange times, $\tau$, for trivalent lanthanides~\cite{Merbach:2005} (points)
scale exponentially with the inverse ionic radius, $1/R$, which is shown as
a proxy for the ion-water interaction strength.
Moving from left to right on the plot, the ions move along the periodic table from Gd$^{3+}$ to Yb$^{3+}$.
For this set of ions, the ion-water interactions are mainly electrostatic in origin.
The solid line is a linear fit to the data points, indicating that $\ln \tau \sim J' \propto 1/R$.
}
\label{fig:exp}
\end{figure}

%
The average exchange and persistence times, $\tau_{\rm X}$ and $\tau_{\rm P}$, respectively,
further support the decoupling of these two processes as $\tilde{T}'$ is lowered, Fig.~\ref{fig:avgfa}.
For large values of $\tilde{T}'$, $\tau_{\rm X}$ and $\tau_{\rm P}$ coincide.
As $\tilde{T}'$ is lowered, $c_{\rm NN}$ decreases, and $\tau_{\rm X}$ and $\tau_{\rm P}$ are increasingly decoupled.
The dynamics controlling exchange and persistence in the solute nearest-neighbor lattice sites
are facilitated by the unconstrained dynamics of the surrounding (bulk) lattice region,
such that
\begin{equation}
\tau_{\rm X}\sim c_{\rm NN}^{-1}
\end{equation}
and
\begin{equation}
\tau_{\rm P}\sim c_{\rm NN}^{-1},
\end{equation}
evidenced by the lines in Fig.~\ref{fig:avgfa}.
Importantly, the linear scaling of $\tau_{\rm X}$ and $\tau_{\rm P}$ with $c_{\rm NN}^{-1}$ is
independent of the dimensionality of the system, which only couples to $c_{\rm B}$; see the Appendix for details.
Note that the scaling of $\tau_{\rm X}$ and $\tau_{\rm P}$ with temperature, $T$,
can be more complex due to the coupling between the bulk and nearest-neighbor sites,
and this scaling also depends on the dimensionality of the system (Appendix).
The data in Fig.~\ref{fig:avgfa} strongly suggests that the exchange and persistence times
scale linearly with $c_{\rm NN}^{-1} = 1+\exp({J' / \kT})$.
The energetic coupling $J'$ captures the net effect of solute-solvent interactions
in the solvation shell,
such that \emph{the water exchange time scales exponentially with the solute-solvent interaction energy}.
Indeed, Lee, Thirumalai, and Hyeon have recently shown that water exchange times scale
exponentially with solute-solvent interactions
within the coordination shell, in addition to an entropic contribution,
in agreement with our predictions~\cite{Lee_2017}.
The energy scale $J'$ is model-dependent, and in classical models typically arises from LJ and Coulomb interactions,
the latter of which ultimately depends on the charge density of the ion.
The scaling with $J'$ is simplified under conditions where only the charge density of the ion is changing,
with remaining interaction potentials held fixed.
In this case, the difference in the energetic coupling between ions arises from electrostatic interactions alone.
Such a situation arises for trivalent lanthanides, for example,
where the major impact of filling the 4f orbitals is an increase in the strength of solute-solvent electrostatic interactions
due to an increase in the charge density (decrease in size) of the ion~\cite{Merbach:2005}.
A Born theory-like model for a hydrated ion would suggest that the electrostatic energy contribution to $J'$ behaves as
$J'\propto Q^2/R$, where $Q$ and $R$ are the charge and radius of the ionic solute, respectively~\cite{Born,Remsing:JPCB:2016}.
Figure~\ref{fig:exp} shows that the experimentally-determined water exchange times around trivalent lanthanides
follow the expected exponential scaling, $\ln \tau \sim J' \propto 1/R$.
Note that we have limited our analysis to ions for which
well-defined exchange times have been determined~\cite{Merbach:2005}.
%

\begin{figure}[tb]
\begin{center}
\includegraphics[width=0.4\textwidth]{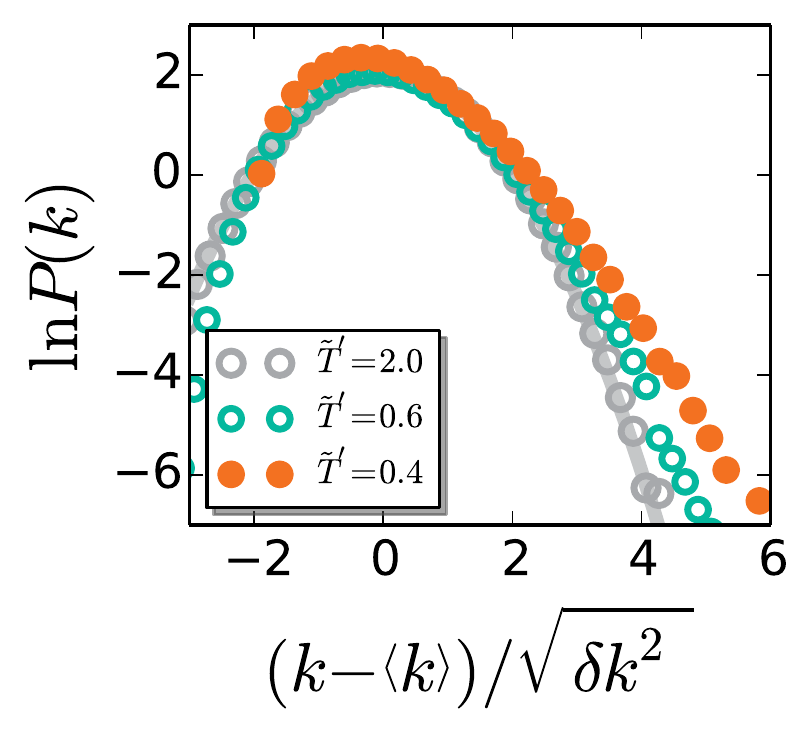}
\end{center}
\caption
{
Probability distributions of the mobility, $P(k)$, for $\tilde{T}'=2.0$, $\tilde{T}'=0.6$, and $\tilde{T}'=0.4$ for the kinetically
constrained model of water exchange, determined for $\Delta t=100$
Monte Carlo sweeps. Points correspond to simulation results and the solid lines
indicates a Gaussian distribution. 
}
\label{fig:mob}
\end{figure}

%
Our results are consistent with the system being in close proximity to a phase transition in trajectory
space~\cite{Chandler_2010,Merolle:PNAS:2005,Hedges_2007}.
This is supported through the examination of probability distributions of trajectory-based observables quantifying the amount
of mobility in a trajectory.
We compute probability distributions of the time-intensive mobility
\begin{equation}
k=\frac{1}{2S\Delta t} \sum_{t=0}^{\Delta t} \sum_{i=1}^{2S} n_i(t),
\end{equation}
where $2S$ is the number of solute neighors ($S=1$ here) and $\Delta t$ is the observation time or trajectory length~\cite{Chandler_2010,Merolle:PNAS:2005}.
At high $\tilde{T}'$, when the persistence and exchange times are coupled and the dynamics are Poissonian,
the distribution $P(k)$ is Gaussian, indicating that the system is not near a spacetime transition, Fig.~\ref{fig:mob}.
In contrast, at low $\tilde{T}'$, when $\tau_{\rm P}$ and $\tau_{\rm X}$ are decoupled, $P(k)$ is significantly
non-Gaussian and displays a fat tail at high mobilities, with increasing non-Gaussian character
as $\tilde{T}'$ is lowered.
This non-Gaussian character indicates the presence of an underlying dynamic phase transition that leads
to the correlated and facilitated dynamics of water exchange.
Similar results have been found for analogous processes involving water bound strongly to extended surfaces~\cite{Willard_2013}.

\section{Conclusions}
We have demonstrated that water exchange reactions exhibit facilitated dynamics
and may be described within the framework of dynamic facilitation theory.
Through the development of a kinetically constrained model, we predict that
water exchange times --- the inverse of the exchange rate --- scale exponentially with the strength of solute-water interactions,
consistent with the ultrasensitivity of water exchange times observed experimentally
and recent empirical findings~\cite{Lee_2017,WaterExchangeReview}.
We conclude by noting that the kinetically constrained models discussed here can be readily extended to study solvent exchange in glassy matrices by imposing kinetic constraints in both the solvent and solute nearest-neighbor regions~\cite{Jack_2006}, with different effective temperatures in the two regions.
Moreover, extensions to two- and three-dimensional lattices will enable the investigation
of the phenomenology underlying directional solvent exchange
and solvent exchange at interfaces,
which are of importance in geochemical~\cite{Casey_2007,AluniteDissolution,WaterRockInteraction,Hofmann_2012,Zarzycki_2015,Remsing:2015ab,Stack_2016,Remsing:CPL:2017}
and electrochemical processes~\cite{Thenuwara_2016,Willard_2013,Limmer_2013,Limmer:CPL:2015,Remsing:2015ab,Remsing:CPL:2017}, for example,
as well as solvent transport through porous solids~\cite{Strong_2016}.
Further extension to multi-site lattice models may also facilitate the study of exchange dynamics
involving ionic solutes and polymeric solvents~\cite{Olender_1,Olender_2}.
We hope that the connections made here between water exchange reactions, kinetically constrained models,
and DF theory will further the development of predictive theories for solvent exchange.
\\

\acknowledgements
This work was supported as part of the
Center for Complex Materials from First Principles (CCM), an Energy Frontier
Research Center funded by the U.S. Department of Energy, Office of Science, Basic Energy
Sciences under Award \#DE-SC0012575.
Computational resources were supported in part by the National Science Foundation
through major research instrumentation grant number 1625061
and by the US Army Research Laboratory under contract number W911NF-16-2-0189.

\appendix
\section{Appendix: Scaling of Exchange Times with Excitation Concentration}
Following previous work~\cite{Jung_2005}, we obtain a mean-field estimate for the behavior of $\tau_{\rm X}$
by noting that the facilitation function for one of the solute nearest neighbor sites is given by
$f_i = n_{i+1}$; the other is $f_i=n_{i-1}$.
This is the same facilitation function as the 1D East model, and we focus on the first neighbor for simplicity.
We start by noting that the rate for the exchange $0\rightarrow 1$ is given by $k^{(+)}=e^{-1/\tilde{T}'} f_i$,
and the rate for the opposite $1\rightarrow0$ process is $k^{(-)}=f_i$.
The mean exchange time for the $n_i=0$ state is then
\begin{equation}
\avg{t_{\rm X}(0)}\approx \frac{1}{\avg{k^{(+)}}} = \frac{e^{1/\tilde{T}'}}{c_{\rm B}},
\end{equation}
and that for the $n_i=1$ state is given by
\begin{equation}
\avg{t_{\rm X}(1)}\approx \frac{1}{\avg{k^{(-)}}} = \frac{1}{c_{\rm B}},
\end{equation}
where we have noted that $\avg{f_i}=c_{\rm B}$.
The total mean exchange time is then given by
\begin{align}
\tau_{\rm X}\equiv\avg{t_{\rm X}}&\approx \frac{1}{2}\brac{\avg{t_{\rm X}(0)} + \avg{t_{\rm X}(1)}} \\
& = \frac{1}{2 c_{\rm NN} c_{\rm B}},
\end{align}
where we have used $1+e^{1/\tilde{T}'}=1/c_{\rm NN}$.
For a general $d$-dimensional square lattice, the average facilitation function is given by~\cite{Jung_2005}
\begin{equation}
\avg{f_i} = 1-(1-c_{\rm B})^{ad},
\label{eq:fid}
\end{equation}
where the product $ad$ indicates the number of sites that can facilitate a change in state.
For the bulk FA and East models, $a=2$ and $a=1$, respectively, independent of $d$.
However, due to the presence of the solute site, $a=a(d)$ is a function of $d$ in our model for solvent exchange.
For example, $a(1)=1$, $a(2)=2/3$, and $a(3)=5/3$ when using the kinetic constraints of the FA model.
We can then carry out the above derivation using Eq.~\ref{eq:fid} to yield
\begin{equation}
\tau_{\rm X} \approx \frac{1}{2 c_{\rm NN}\brac{1-(1-c_{\rm B})^{ad}}},
\end{equation}
which illustrates that $\tau_{\rm X}\sim c_{\rm NN}^{-1}$ for all $d$, such that the scaling with
solute-solvent interactions is also independent of $d$.
Note that the temperature dependence of $\tau_{\rm X}$ is more complex because
both $c_{\rm NN}$ and $c_{\rm B}$ depend on $T$.
The mean persistence time is related to the moments of the exchange time
through~\cite{Jung_2005}
\begin{equation}
\tau_{\rm P} = \frac{\avg{t_{\rm X}^2}}{2\avg{t_{\rm X}}}.
\end{equation}
Using the variance, this can be rewritten as
\begin{equation}
\tau_{\rm P} = \frac{\avg{t_{\rm X}}}{2} \brac{ 1 - \frac{ \avg{\delta t_{\rm X}^2} } {\avg{t_{\rm X}}^2} } \sim c_{\rm NN}^{-1},
\end{equation}
where $\delta t_{\rm X}\equiv t_{\rm X}-\avg{t_{\rm X}}$.

\end{document}